\documentclass[12pt]{article}
\usepackage{amsmath}
\usepackage{latexsym}
\usepackage{amssymb, amsmath, xspace, lscape,  latexsym,amsthm}
\usepackage{amsfonts}
\usepackage{mathrsfs}
\usepackage{epsf}

\newcommand{\bea}{\begin{eqnarray}}
\newcommand{\eea}{\end{eqnarray}}
\def\beq#1#2\eeq{
        \begin{equation}
        \label{#1}
            #2
        \end{equation}}

\newcommand{\vp}{\textsf{v}}
\newcommand{\px}{\partial_x}
\newcommand{\py}{\partial_y}

\newcommand{\pa}{\partial_a}
\newcommand{\pb}{\partial_b}
\newcommand{\tH}{\tilde{H}}

\newcommand{\al}{\alpha}

\newcommand{\bt}{\beta}

\newcommand{\re}{\textrm{e}}
\newcommand{\tp}{\textsf{p}_1}

\renewcommand{\tilde}{\widetilde}

\def\btheor#1\etheor{
        \begin{theor}
            #1
        \end{theor}
    }

    \def\bsled#1\esled{
        \begin{sled}
            #1
        \end{sled}   }

\newtheorem{lemma}{Lemma}

\def\hm#1{#1\nobreak\discretionary{}{\hbox{\m@th$#1$}}{}}
\def\mi#1{\discretionary{\hbox{\m@th$#1$}}{\hbox{\m@th$#1$}}{}}

\vspace{4ex}
\begin{document}
\title{\bf PDEs satisfied by extreme eigenvalues distributions of
GUE and LUE}
\author{
\quad Estelle Basor\footnotemark[1],\quad Yang Chen\footnotemark[2],\quad Lun Zhang\footnotemark[3]}
\date{\today}
\maketitle
\renewcommand{\thefootnote}{\fnsymbol{footnote}}
\footnotetext[1]{ American Institute of Mathematics, Palo Alto,
California 94306, USA, E-mail address: ebasor@aimath.org.}
\renewcommand{\thefootnote}{\fnsymbol{footnote}}
\footnotetext[2]{Department of Mathematics, Imperial College London,
180 Queen's Gate, London SW7 2BZ, UK. E-mail address:
y.chen@imperial.ac.uk.}
\renewcommand{\thefootnote}{\fnsymbol{footnote}}
\footnotetext[3]{Department of Mathematics, Katholieke Universiteit
Leuven, Celestijnenlaan 200B, B-3001 Leuven, Belgium. E-mail
address: lun.zhang@wis.kuleuven.be.}

\begin{abstract}
In this paper we study, $\textsf{Prob}(n,a,b),$ the probability
that all the eigenvalues of finite $n$ unitary ensembles lie in the interval
$(a,b)$. This is identical to the probability that the largest
eigenvalue is less than $b$ and the smallest eigenvalue is greater
than $a$. It is shown that a quantity allied to
$\textsf{Prob}(n,a,b)$, namely,
$$
H_n(a,b):=\left[\frac{\partial}{\partial a}+\frac{\partial}{\partial
b}\right]\ln\textsf{Prob}(n,a,b),
$$
in the Gaussian Unitary Ensemble (GUE) and
\begin{equation*}
H_n(a,b):=\left[a\frac{\partial}{\partial a}+b\frac{\partial
}{\partial b}\right]\ln \textsf{Prob}(n,a,b),
\end{equation*}
in the Laguerre Unitary Ensemble (LUE) satisfy certain nonlinear
partial differential equations for fixed $n$, interpreting
$H_n(a,b)$ as a function of $a$ and $b$. These partial differential
equations maybe considered as two variable generalizations of a
Painlev\'{e} IV and a Painlev\'{e} V system, respectively. As an
application of our result, we give an analytic proof that the
extreme eigenvalues of the GUE and the LUE, when suitably centered and scaled,
are asymptotically independent.


\end{abstract}
\noindent


\setcounter{equation}{0}
\section{Introduction}
In the theory of random matrices, the study of eigenvalue
distribution attracts the most interest and has many applications in
both mathematic, physics and wireless communications; see for
example \cite{chen+bai,chen+mckay,Deift,Mehta}. It is by now a
classic result that the largest eigenvalue distribution of the
Gaussian Unitary Ensemble (GUE) and Laguerre Unitary Ensemble (LUE),
the celebrated Tracy-Widom II distribution \cite{twa}, denoted as
$F_2$, is given by a integral involving the Hastings-McLeod solution
of the Painlev\'e~II equation. For the GUE, it can be immediately
seen via a change of variables that the analogous result holds for
the smallest eigenvalue. These distributions emerge after centering
the extreme eigenvalues at $\pm \sqrt{2n}$, the edges of the GUE
spectrum, followed by a scaling with respect to the density at the
soft edge; see \cite{twa} for the original derivation of this
particular $P_{II}$. In \cite{twdet} a finite $n$ version of the
result was found, which turns out to be a $P_{IV}$, with the
limiting $F_2$ after the above centering and scaling of the largest
eigenvalues. For LUE, the distribution is obtained after centering
the largest eigenvalues at $4n$ and a scaling at the soft edge of
the LUE spectrum. The finite $n$ analogue in this case was found to
be a $P_{V}$ \cite{twdet}.

In this paper we consider a related problem. We are concerned with
the probability that all the eigenvalues are in an interval $(a,b)$.
This is of course equivalent to the probability that the largest
eigenvalue is less than $b$ and the smallest one greater than $a$.
It turns out that this probability is related to another expression
that is the solution of a nonlinear partial differential equation
(pde). In the GUE case, this pde maybe interpreted as a two variable
version of a Painlev\'{e} IV $\sigma$-form, since in the limit $a$
fixed, $b\to \infty$ or $b$ fixed, $a\to -\infty$, the pde reduces
to the ode corresponding to the ``left'' or ``right'' edge version
of $P_{IV}$. We have a similar interpretation for the pde derived in
the LUE case as a two variable version of a Painlev\'{e} V
$\sigma$-form. See also \cite{van} where the authors derived pdes
for the logarithm of this probability which are related to KP
equations using vertex operators and the associated Virasoro
constraint \cite{Adl+Van}.

Our method is based on a ladder operators formalism for orthogonal
polynomials \cite{chen1} and the associated compatibility
conditions. This is, by now, a well-known method that has been
applied to obtain exact solutions in a classical case \cite{chen2}
and adapted to orthogonal polynomials on the unit circle
\cite{chen3} intimately related to the theory of Toeplitz matrices.
One may find in \cite{chen+its} a list of references on this
formalism. Recent examples on the applications of the ladder
operators with the resulting Painlev\'e equations can be found in
\cite{chen+basor,chen+dai,chen+feigin,chen+pruessner,chen+zhang,Dai+Zhang,Forrester+Ormerod};
see also \cite{chen+mckay} for an application to the information
theory of multiple input and multiple output wireless communications
which involves certain deformation of the LUEs. In particular, a
comparison between the ladder operator theory and the isomonodromy
theory of Jimbo-Miwa-Ujimo \cite{JMU} is carried out in
\cite{chen+its} and \cite{Forrester+Ormerod} for different specific
Hermitian random ensembles. The extension of ladder operators to
discrete orthogonal polynomials and $q$-orthogonal polynomials is
given in \cite{INS} and \cite{chen+ismail08}, respectively. From
these extensions, it is shown that the recurrence coefficients of
certain discrete or $q$-orthogonal polynomials are related to the
discrete or $q$-Painlev\'{e} equations, respectively. We refer to
\cite{B thesis,BFV,Boelen+walter,IM} for investigations of this
aspect.

Recently, based entirely on the estimate on the integral operator in
a Fredholm expansion, the authors of \cite{bianchi} showed that the
extreme eigenvalues of GUE, when suitably centered and scaled, are
asymptotically independent random variables. As an application of
our result, we give a proof that is an analytic counter-part of this
probabilistic version. Our strategy is to scale the two variables in
the pde obtained near both edges of the GUE spectrum, which will
lead to a limiting pde. It turns out that the solution to the
limiting pde has a particular form which implies that the joint
probability density is a product of two independent densities, that
is, there is asymptotic independence. With different and delicate
scalings, we show that the solution of limiting pde in LUE case is
also asymptotically equal to the sum of Tracy-Widom left and right
distributions.

The rest of this paper is organized as follows. In section 2, we
give a summary of the ladder operators method, the associated
compatibility conditions and a summation identity. From these
conditions, we derive a system of non-linear difference equations
satisfied by auxiliary variables that appear naturally in this
approach. These difference equations will be instrumental in our
further derivation of the pdes. We study GUE in section \ref{study
of GUE} and LUE in section \ref{study of LUE}. As an application of
our result, we give an analytic proof that the extreme eigenvalues
of GUE and LUE, when suitably centered and scaled, are asymptotic
independent.

\setcounter{equation}{0}
\section{Preliminaries}\label{sec:preli}

In the theory of Hermitian random matrices, one encounters the
following (un-normalized) joint probability density of eigenvalues
 $\{x_j\}_{j=1}^{n}$:
\begin{equation}
 p(x_1,...,x_n)  =  \prod_{1\leq i<j\leq n}[\Delta_n(x)]^{2}
\prod_{k=1}^{n}w(x_k),
\end{equation}
where
\begin{equation}\label{deltan}
\Delta_n(x):=\prod_{1\leq j<i\leq n}(x_i-x_j),
\end{equation}
and $w$ is a weight function defined on an interval $I\subseteq
\mathbb{R}$. The GUE corresponds to $w(x)=\re^{-x^2}$ and
$I=\mathbb{R}$, while the LUE corresponds to
$w(x)=x^{\alpha}\re^{-x}$, $\alpha>0$ and $I=(0,\infty)$. In what
follows everything depends on $n$, but the dependence will only be
displayed when it is easy to do so or needed for clarification.

Denote by $\textsf{Prob}(n,a,b)$ the probability that all
eigenvalues lie in an interval $(a,b) \subset I$. We then have
\begin{equation}\label{probability}
\textsf{Prob}(n, a, b)=\frac{\int_{(a,b)^{n}}[\Delta_n(x)]^{2}
\prod_{k=1}^{n}w(x_k)dx_k}{\int_{I^{n}}
[\Delta_n(x)]^{2}\prod_{k=1}^{n}w(x_k)dx_k}.
\end{equation}
It is a well-known fact (cf. \cite{Szego}) that the multiple
integrals in \eqref{probability} can be expressed as determinant of
Hankel matrix generated by the weight function $w$. The moments of
the weight are defined by
\begin{equation}
\mu_{j}(a,b):=\int_{a}^{b}x^j w(x)dx, \qquad j=0,1,...,
\end{equation}
and the Hankel determinant is the determinant of the moment matrix
$(\mu_{j+k})_{j,k=0}^{n-1}$:
\begin{align}\label{Hankel det}
D_n(a,b)&:=\det\left(\mu_{j+k}(a,b)\right)_{j,k=0}^{n-1}\nonumber\\
&=\prod_{j=0}^{n-1}h_j(a,b)\nonumber\\
&=
\frac{1}{n!}\int_{(a,b)^{n}}[\Delta_n(x)]^{2}\prod_{k=1}^{n}w(x_k)dx_k.
\end{align}
Here, $h_i(a,b)$ is the square of the $L^2$ norm of the (monic)
polynomials orthogonal with respect to $w$ over $(a,b)$, i.e.,
\begin{equation}\label{def:pn}
\int_{a}^{b}P_i(x,a,b)P_j(x,a,b)w(x)dx=h_i(a,b) \delta_{i,j}.
\end{equation}
The monic polynomials $P_n(z,a,b)$ are normalized as
\begin{equation}
P_n(z,a,b)=z^n+\tp(n,a,b)z^{n-1}+...+P_n(0,a,b).
\end{equation}
Therefore,
\begin{equation}
\textsf{Prob}(n,a,b)=\frac{D_n(a,b)}{D_n(I)}
=\prod_{j=0}^{n-1}\frac{h_j(a,b)}{h_j(I)}.
\end{equation}
It is then clear that in order to compute the probability
$\textsf{Prob}(n,a,b)$, we need to compute the product of the norms
$h_{i}(a,b)$ and thus we need information about the orthogonal
polynomials. According to the general theory of orthogonal
polynomials, an immediate consequence of \eqref{def:pn} is the three
terms recurrence relations:
\begin{equation}
zP_{n}(z,a,b)=P_{n+1}(z,a,b)+\al_{n}(a,b)P_n(z,a,b)+\bt_{n}(a,b)P_{n-1}(z,a,b)
\end{equation}
with the initial conditions
\begin{equation}
P_0(z)=1,\qquad \bt_0P_{-1}(z)=0.
\end{equation}
An easy consequence of the recurrence relation is
\begin{equation}
\label{alphan representation}\al_{n}(a,b)=\tp(n,a,b)-\tp(n+1,a,b).
\end{equation}

In the next two sections, we will give an account for a recursive
algorithm for the determination of the recurrence coefficients
$\al_n$, $\bt_n$ in the GUE  and LUE, respectively, based on a pair
of ladder operators and the associated supplementary conditions. It
will become clear that the determination of $\al_{n}$ and $\bt_{n}$
will yield the necessary information to find conditions on
$h_{n}(a,b)$. The following three lemmas can be found in, for
example, \cite{chen+its} and the extensive references therein. For
convenience, we suppress the $a$, $b$ dependence in $\alpha_{n}$,
$\beta_{n}$ and $h_{n}$ in these lemmas.

\begin{lemma}\label{Lem:Ladder relation}
Suppose that $w$ is such that the moments \bea
\mu_{i}(a,b):=\int_{a}^{b}x^i w(x) dx,\quad i=0,1,... \eea exist and
that $\vp=-\ln w$ has a derivative in some Lipshitz class with
positive exponent. The lowering and raising operators satisfy the
following: \begin{align}
P_n'(z,a,b)&=-B_n(z,a,b)P_n(z,a,b)+\bt_{n}A_n(z,a,b)P_{n-1}(z,a,b),\\
P_{n-1}'(z,a,b)&=[B_n(z,t)+\vp'(z)]P_{n-1}(z,a,b)-A_{n-1}(z,a,b)P_n(z,a,b),
\end{align} where
\[
A_n(z,a,b) :=
\frac{w(y)P_n^2(y,a,b)}{h_n(y-z)}\bigg|_{a}^{b}+\frac{1}{h_n}\int_{a}^{b}
\frac{\vp'(z)-\vp'(y)}{z-y} P_n^2(y,a,b)w(y)dy, \]
\[B_n(z,a,b) := \frac{w(y)P_{n}(y,a,b)P_{n-1}(y,a,b)}{h_{n-1}(y-z)}\bigg|_{a}^{b}
+
\frac{1}{h_{n-1}}\int_{a}^{b}\frac{\vp'(z)-\vp'(y)}{z-y}
P_n(y,a,b)P_{n-1}(y,a,b)w(y)dy.\]
\end{lemma}
A direct calculation produces two fundamental supplementary (compatibility) conditions valid
for all $z$ and these are stated in two different forms in the next lemma.
\begin{lemma}
 The functions $A_{n}(z,a,b)$ and  $B_{n}(z,a,b)$ satisfy the conditions:
$$B_{n+1}(z,a,b)+B_n(z,a,b)=(z-\al_n)A_n(z,a,b)-\vp'(z), \eqno(S_1)
$$
$$
1+(z-\al_n)(B_{n+1}(z,a,b)-B_n(z,a,b))=\bt_{n+1}A_{n+1}(z,a,b)-\bt_nA_{n-1}(z,a,b).
\eqno(S_2)
$$
\end{lemma}
It turns out that there is an equation which gives better insight into the
$\al_n$ and $\bt_n$ if $(S_1)$ and $(S_2)$ are suitably combined.
\begin{lemma}
The functions $A_{n}(z,a,b)$, $B_{n}(z,a,b)$ and
$$\sum_{j=0}^{n-1}A_j(z,a,b)$$
satisfy the condition:
$$
B^2_n(z,a,b)+\vp'(z)B_n(z,a,b)+\sum_{j=0}^{n-1}A_j(z,a,b)=\bt_nA_n(z,a,b)A_{n-1}(z,a,b).
\eqno(S'_2).
$$
\end{lemma}

\setcounter{equation}{0}
\section{Studies of Gaussian Unitary Ensembles}\label{study of GUE}
It is the aim of this section to show that a quantity associated
with $\textsf{Prob}(n,a,b)$ defined in \eqref{probability} satisfies
a nonlinear pde for GUE via the ladder operators introduced in the
previous section. To this end, we recall that in the present case
$w(x)=\re^{-x^2}$ and $I=\mathbb{R}$. Hence, an appeal to lemma 1
gives
\begin{align}
A_n(z,a,b)&=\frac{R_{n,b}}{z-b} + \frac{R_{n,a}}{z-a}+2, \\
B_n(z,a,b)&=\frac{r_{n,b}}{z-b} +\frac{r_{n,a}}{z-a},
\end{align}
where
\begin{align}
R_{n,b} &= -\re^{-b^2}\frac{P^2_n(b,a,b)}{h_n(a,b)},\\
R_{n,a} &= \re^{-a^2}\frac{P^2_n(a,a,b)}{h_n(a,b)},\\
r_{n,b}&=-\re^{-b^2}\frac{P_n(b,a,b)P_{n-1}(b,a,b)}{h_{n-1}(a,b)},\\
r_{n,a}&=\re^{-a^2}\frac{P_n(a,a,b)P_{n-1}(a,a,b)}{h_{n-1}(a,b)}.
\end{align}
Substituting these into $(S_1)$ implies
\begin{align}
R_{n,a}+R_{n,b}&= 2\al_{n}, \label{Rna+Rnb}\\
r_{n+1,b} +r_{n,b}&= (b - \al_{n})R_{n,b},\\
r_{n+1,a} +r_{n,a}&= (a - \al_{n})R_{n,a};
\end{align}
while into $(S_2')$,
\begin{align}
\bt_n&=\frac{n}{2}+\frac{r_{n,a}}{2} + \frac{r_{n,b}}{2}, \label{beta-r}\\
r_{n,a}^2&=\bt_nR_{n,a}R_{n-1,a}, \label{rnaR}\\
r_{n,b}^2&=\bt_nR_{n,b}R_{n-1,b},\label{rnbR}\\
2\frac{r_{n,b}r_{n,a}}{b-a}
+2br_{n,b}+\sum_{j=0}^{n-1}R_{j,b}&=\bt_n\biggl[\frac{R_{n,b}R_{n-1,a}
+R_{n-1,b}R_{n,a}}{b-a}
 +2(R_{n-1,b}+R_{n,b})\biggr],\\
2\frac{r_{n,b}r_{n,a}}{a-b} +2ar_{n,a}+\sum_{j=0}^{n-1}R_{j,a}&=
\bt_n\biggl[\frac{R_{n,a}R_{n-1,b}+R_{n-1,a}R_{n,b}}{a-b}
+2(R_{n-1,a}+R_{n,a})\biggr].
\end{align}
The sum of the last two equations gives
\begin{align}
 2br_{n,b}+2ar_{n,a} + \sum_{j=0}^{n-1}(R_{j,a}+R_{j,b})
 &=2\bt_n( R_{n-1,b}+R_{n,b}+R_{n-1,a}+R_{n,a})\nonumber\\
 &=2\bt_n(R_{n,a}+R_{n,b})+ 2\frac{r_{n,a}^2}{R_{n,a}}+2\frac{r_{n,b}^2}{R_{n,b}},
 \label{rrRR}
\end{align}
where we have made use of \eqref{rnaR} and \eqref{rnbR} to eliminate
the terms $R_{n-1,a}$, $R_{n-1,b}$ and arrive at the last equation.

\subsection{Toda evolution}
Before coming to the derivation of pde, we first establish some
connections between $\beta_n$, $\tp(n,a,b)$ and the auxiliary
constants $r_{n,a}$, $r_{n,b}$, $R_{n,a}$ and $R_{n,b}$, which will
lead to a two variable analogue of the Toda equations for the
recurrence coefficients $\alpha_n$ and $\beta_n$. These relations
will be helpful in our further analysis.

We start with taking the partial derivative of $h_n$ with respect to
$b$ in \eqref{def:pn}, which gives
\begin{equation}
\partial_bh_n = \re^{-b^2}P^2_n(b,a,b),
\end{equation}
and consequently
\begin{equation}
\pb\ln h_n= -R_{n,b}.
\end{equation}
Since $\bt_n=h_{n}/h_{n-1}$, we find,
\begin{equation}
\frac{1}{\bt_n}\pb\bt_n =R_{n-1,b} -R_{n,b},\label{beta-b}
\end{equation}
and a similar computation yields
\begin{equation}
\frac{1}{\bt_n}\pa\bt_n = R_{n-1,a} -R_{n,a}.\label{beta-a}
\end{equation}
Again taking partial derivatives with respect to $b$ on both sides
of the equation
$$
0=\int_{a}^{b}P_{n}(x,a,b)P_{n-1}(x,a,b)\re^{-x^2}dx
$$
produces
\begin{align}
0&= P_{n}(b,a,b)P_{n-1}(b,a,b)\re^{-b^2}+
\int_{a}^{b}\left[\pb\textsf{p}_1(n,a,b) x^{n-1}+\cdots
\right]P_{n-1}(x,a,b)
\re^{-x^2}dx\nonumber\\
&=P_{n}(b,a,b)P_{n-1}(b,a,b)\re^{-b^2}+\pb\textsf{p}_1(n,a,b)h_{n-1}(a,b),\nonumber
\end{align}
and consequently
\begin{equation}\label{pa p1}
\pb\tp(n,a,b)=
-\re^{-b^2}\frac{P_n(b,a,b)P_{n-1}(b,a,b)}{h_{n-1}(a,b)}=r_{n,b}.
\end{equation}
A similar computation yields
\begin{equation}\label{pb p1}
\pa\tp(n,a,b)=
\re^{-a^2}\frac{P_n(a,a,b)P_{n-1}(a,a,b)}{h_{n-1}(a,b)}= r_{n,a}.
\end{equation}
Now, we are ready to prove:
\begin{lemma}
For GUE, we have
\begin{align}\label{betan in alphan GUE}
\frac{1}{\bt_n}(\pa+\pb)\bt_n&=2(\al_{n-1}-\al_n), \\
\label{alphan in betan GUE} (\pa+\pb)\al_n&=2(\bt_{n}-\bt_{n+1})-1.
\end{align}
\end{lemma}
\begin{proof}
Taking a sum of \eqref{beta-b} and \eqref{beta-a}, the equation
\eqref{betan in alphan GUE} is immediate from \eqref{Rna+Rnb}. To
estblish \eqref{alphan in betan GUE}, we note from the sum of
\eqref{pa p1} and \eqref{pb p1} that
\begin{equation}
(\pa+\pb)\textsf{p}_1(n,a,b)=r_{n,a}+r_{n,b}=2\bt_n-n,\nonumber
\end{equation}
where the last equality follows from \eqref{beta-r}. This, together
with \eqref{alphan representation}, gives us \eqref{alphan in betan
GUE}.
\end{proof}
The equations \eqref{betan in alphan GUE} and \eqref{alphan in betan
GUE} constitute a two variable version of the Toda equations.

\subsection{Derivation of partial differential equation}
\label{sec:pde in GUE} To this end, we set
\begin{equation}\label{Hn GUE}
H_n(a,b):=(\pa+\pb)\ln D_n(a,b),
\end{equation}
where $D_n$ is the Hankel determinant defined in \eqref{Hankel det}
associated with GUE. It is the aim of this section to derive a pde
satisfied by $H_n$. Our strategy is to construct a system of partial
differential equations in the functions $r_{n,a}$, $r_{n,b}$,
$R_{n,b}$ and $R_{n, a}$, because they provide a direct link to
$H_n$.

To see this, note that
\begin{equation}
\ln D_{n}(a,b) = \sum_{j=0}^{n-1} \ln h_{j}(a,b),
\end{equation}
and thus \bea \label{part a ln Dn} \pa\ln D_{n}(a,b) =
\sum_{j=0}^{n-1}
\pa\ln h_{j}(a,b) = -\sum_{j=0}^{n-1}R_{j,a}, \\
\label{part b ln Dn} \pb\ln D_{n}(a,b)= \sum_{j=0}^{n-1} \pb\ln
h_{j}(a,b) = -\sum_{j=0}^{n-1}R_{j,b}. \eea
Recall that we have from
\eqref{Rna+Rnb}
\bea
 R_{n,b} + R_{n,a}= 2\al_{n},\nonumber
\eea and that from \eqref{alphan representation} \bea
\al_{n}(a,b)=\tp(n,a,b)-\tp(n+1,a,b).\nonumber
 \eea
Therefore we have that
\bea
\sum_{j=0}^{n-1}\left( R_{j,b}+R_{j,a}\right)
 = 2
\sum_{j=0}^{n-1}\al_{j}(a,b) = - 2\tp(n,a,b). \label{Rp} \eea A
combination of \eqref{Hn GUE}, \eqref{part a ln Dn} and \eqref{part
b ln Dn} gives us
\begin{align}\label{Hn rep}
H_n&=(\partial_a+\partial_b)\ln
D_n=-\sum_{j=0}^{n-1}(R_{j,a}+R_{j,b})
=2\tp(n,a,b)\nonumber\\
&=2ar_{n,a}+2br_{n,b}-2\frac{r_{n,a}^2}{R_{n,a}}
-2\frac{r_{n,b}^2}{R_{n,b}}-2\bt_n(R_{n,a}+R_{n,b}),
\end{align}
where the second equality of \eqref{Hn rep} follows from
\eqref{rrRR}. In view of \eqref{pa p1} and \eqref{pb p1}, we note
that \bea
\partial_a H_n=2r_{n,a}, \qquad \partial_b H_n= 2r_{n,b}. \label{Hn}
\eea
From \eqref{beta-a}, \eqref{beta-b}, \eqref{rnaR} and
\eqref{rnbR}, we have
\begin{align}
\pa\bt_n&=\frac{r_{n,a}^2}{R_{n,a}}-\bt_{n}R_{n,a},
\\
\pb\bt_n&=\frac{r_{n,b}^2}{R_{n,b}}-\bt_{n}R_{n,b}.
\end{align}
This, together with \eqref{beta-r}, implies
\begin{align*}
\frac{1}{2}\pa(r_{n,a}+r_{n,b})&=\frac{r_{n,a}^2}{ R_{n,a}}-\Big(\frac{n}{2}+\frac{r_{n,a}+r_{n,b}}{2}\Big)R_{n,a},\\
\frac{1}{2}\pb(r_{n,a}+r_{n,b})&=\frac{r_{n,b}^2}{
R_{n,b}}-\Big(\frac{n}{2}+\frac{r_{n,a}+r_{n,b}}{2}\Big)R_{n,b}.
\end{align*}

Expressing $r_{n,a}$ and $r_{n,a}$ in terms of the partial
derivatives of $H_n$, we have
 \begin{align}
\partial_a^2H_n+\partial_a\partial_bH_n&=\frac{(\partial_a H_n)^2}{R_{n,a}}-\big(2n+\partial_a H_n+\partial_b H_n\big)R_{n,a},
\label{part-1} \\
\partial_b^2H_n+\partial_a\partial_bH_n&=\frac{(\partial_b H_n)^2}{R_{n,b}}-\big(2n+\partial_a H_n+\partial_b H_n\big)R_{n,b}.
 \label{part-2}
 \end{align}
We may consider \eqref{part-1} and \eqref{part-2} as quadratic
equations in $R_{n,a}$ and $R_{n,b}.$ Solving for them and
substituting into \eqref{Hn rep}, we find, after some
simplification,
%
 \begin{align}\label{pde in GUE}
2b\partial_b H_n+2a\partial_a
H_n-2H_n=& \sqrt{(\partial_a^2H_n+\partial_a\partial_bH_n)^2+
4(\partial_a H_n)^2(2n+\partial_a H_n+\partial_b H_n)}\nonumber\\
& -\sqrt{(\partial_b^2H_n+\partial_a\partial_bH_n)^2+4(\partial_b
H_n)^2(2n+\partial_a H_n+\partial_b H_n)}.
\end{align}
(Note that in the above the signs of the square roots are determined by the signs of $R_{n,a}$ and $R_{n,b}$. The former is positive and the latter is negative and the term $2n+\partial_a H_n+\partial_b H_n$ is also positive since it is the same as $4\beta_{n}$.)
After clearing the square roots, we obtain that the function $H_n$
defined in \eqref{Hn GUE} satisfies the following pde:
\begin{align}\label{pde in GUE full}
\big(&(2b\,\partial_b H_n+2a\,\partial_a H_n-2H_n)^{2} -
((\partial_a^2H_n+\partial_a\partial_bH_n)^2+ 4\big(\partial_a
H_n\big)^2 (2n+\partial_a H_n+\partial_b H_n )) \nonumber \\
&~~~~~~~~~~~~~~~~~~~~~-
((\partial_b^2H_n+\partial_a\partial_bH_n\big)^2+4\big(\partial_b
H_n)^2
(2n+\partial_a H_n+\partial_b H_n))\big)^{2}  \nonumber \\
 &=
 4\big((\partial_a^2H_n+\partial_a\partial_bH_n)^2+
4(\partial_a H_n)^2(2n+\partial_a H_n+\partial_b H_n)\big) \nonumber \\
 &~~~~~~~~~~~~~~~~~~~~~\times\big((\partial_b^2H_n+\partial_a\partial_bH_n)^2+4(\partial_b H_n)^2
(2n+\partial_a H_n+\partial_b H_n)\big).
\end{align}

In our approach, the end points $a$ and $b$ are the ``times'',
although they play a distinct role from those in a two variable
generalization of Painlev\'e IV \cite{tsuda}.

Suppose $H_n$ is independent of $a$, the equation \eqref{pde in GUE
full} reduce to
\begin{equation}
(\partial_b^2 H_n)^2=4(b\partial_b H_n-H_n)^2-4(\partial_b
H_n)^2(2n+\partial_b H_n),
\end{equation}
which actually is the Okamoto-Jimbo-Miwa $\sigma$-form of the
Painlev\'{e} IV equation \cite{JM}:
\begin{equation}
(\sigma'')^2=4(z\sigma'-\sigma)^2-4(\sigma'+\nu_0)(\sigma'+\nu_1)(\sigma'+\nu_2)
\end{equation}
with
\begin{equation}
\nu_0=2n, \qquad \nu_1=\nu_2=0.
\end{equation}
The same conclusion holds if $H_n$ is independent of $b$.

Finally, it may be obvious, but worth pointing out that the solution
$H_{n}$ to the pde along with initial conditions does indeed yield
the desired probability. This is because
\begin{equation}
\ln \textsf{Prob}(n, a, b) = \int_{0}^{a} H_{n}(t, t + b - a) dt  +
\ln \textsf{Prob}(n, 0, b-a).
\end{equation}

\setcounter{equation}{0} \subsection{Asymptotic independence of the
extreme eigenvalues in GUE} \label{asy ind}

As an application of the pde derived in section \ref{sec:pde in
GUE}, we show in this section that the extreme eigenvalues of GUE,
when suitably centered and scaled, are asymptotic independent, i.e.,
\begin{equation}\label{asy independent}
\begin{aligned}
&\lim_{n\to\infty}\textsf{Prob}
\left(n,\;\left(\lambda_{\textsf{min}}+\sqrt{2n}\right)n^{1/6}/c>x,\left(\sqrt{2n}-\lambda_{\textsf{max}}\right)n^{1/6}/c<y
\right)
\\
&=\lim_{n\to\infty}\textsf{Prob}
\left(\left(\lambda_{\textsf{min}}+\sqrt{2n}\right)n^{1/6}/c>x\right)\lim_{n\to\infty}\textsf{Prob}
\left(\left(\sqrt{2n}-\lambda_{\textsf{max}}\right)n^{1/6}/c<y\right),
\end{aligned}
\end{equation}
where $\lambda_{\textsf{min}}$ ($\lambda_{\textsf{max}}$) denotes
the smallest (largest) eigenvalue.

Our method is to scale $a$ and $b$ near the edges of the spectrum
and compute asymptotically the resulting pde. For this purpose, we
let
\begin{equation}\label{change of variable}
a=-\sqrt{2n}+ c \frac{x}{n^{1/6}}, \quad  b=\sqrt{2n}-c
\frac{y}{n^{1/6}},
\end{equation}
with $c>0$, and note that
$$\pa=\frac{n^{1/6}}{c}\partial_x,\qquad \pb=-\frac{n^{1/6}}{c}\partial_y.$$
Recall
\begin{equation} H_n(a,b)=(\pa+\pb)\ln
D_n(a,b)=(\pa+\pb)\ln\textsf{Prob}(n,a,b),\nonumber
\end{equation} and in the
$x$, $y$ variables this becomes \bea
\frac{c}{n^{1/6}}H_n\left(-\sqrt{2n}+c\frac{x}{n^{1/6}},\sqrt{2n}-c
\frac{y}{n^{1/6}}\right) =(\partial_x-\partial_y)\ln D_n. \eea Let
\bea
\tH(x,y,n):=\frac{c}{n^{1/6}}H_n\left(-\sqrt{2n}+c\frac{x}{n^{1/6}},\sqrt{2n}-c\frac{y}{n^{1/6}}\right).
\eea After substituting the change of variables \eqref{change of
variable}, the leading term of \eqref{pde in GUE} is of order
$n^{4/3}$ and produces the limiting pde
\begin{align}\label{limiting pde}
&-8\sqrt{2}c^3\tH\py\tH\px\tH+8\sqrt{2}c^3y(\py\tH)^2\px\tH-4(\py\tH)^3\px\tH+\px\tH(\py^2\tH-\px\py\tH)^2
\nonumber\\
&+\py\tH(8\sqrt{2}c^3x(\px\tH)^2+4(\px\tH)^3+(\px\py\tH-\px^2\tH)^2)=0.
\end{align}
To ascertain whether the scaled smallest and largest eigenvalues may
be described by their respective Tracy-Widom law for the extreme
eigenvalues, we make use of a factorization ansatz, \bea
\tH(x,y)=f(x)+g(y), \eea where $f(x)$ and $g(y)$ satisfy the
$\sigma$-form of a particular Painlev\'e II. That is,
\begin{align}
\frac{1}{4}(f''(x))^2&=2\sqrt{2}c^3f(x)f'(x)-2\sqrt{2}c^3x(f'(x))^2-(f'(x))^3,\\
\frac{1}{4}(g''(y))^2&=2\sqrt{2}c^3g(y)g'(y)-2\sqrt{2}c^3y(g'(y))^2+(g'(y))^3.
\end{align}
We use this ansatz because if $x=-\infty$ or $y=\infty,$ then the
functions $f$ and $g$ yield the correct solution of the pde and we
believe that the solution should be their sum. Indeed this is true.
An simple computation shows that \eqref{limiting pde} is satisfied
identically.

Now, we set \bea \textsf{P}(x,y)=\lim_{n\to\infty}\textsf{Prob}
\left(n,\;\left(\lambda_{\textsf{min}}+\sqrt{2n}\right)n^{1/6}/c>x,\left(\sqrt{2n}-\lambda_{\textsf{max}}\right)n^{1/6}/c<y
\right).\eea  Recall that \bea \left(\px-\py\right)\ln
\textsf{P}(x,y)=\lim_{n\to\infty}\tH(x,y,n). \eea Therefore the
general solution of $\textsf{P}(x,y)$ is of this form \bea
\textsf{P}(x,y)=F(x)G(y)\exp\left[\Psi(x+y)\right], \eea where
\begin{align}
F(x)&=\lim_{n\to\infty}\textsf{Prob}
\left(\left(\lambda_{\textsf{min}}+\sqrt{2n}\right)n^{1/6}/c>x\right),\\
G(y)&=\lim_{n\to\infty}\textsf{Prob}
\left(\left(\sqrt{2n}-\lambda_{\textsf{max}}\right)n^{1/6}/c<y\right),
\end{align}
and $\Psi$ is an arbitrary $C^1$ function. In view of \eqref{asy
independent}, it remains to show $\Psi\equiv 0$. To see this, note
that
\begin{align}
&\lim_{x\to -\infty}F(x)=1,\quad\lim_{x\to\infty}F(x)=0,\\
&\lim_{y\to -\infty}G(y)=0,\quad\lim_{y\to\infty}G(y)=1,
\end{align} and
\bea \lim_{x\to -\infty,y\to\infty}\textsf{P}(x,y)=1. \eea Take a
fixed $z$ and let $x+y=z$. We see that \bea \lim_{x\to
-\infty}\textsf{P}(x,z-x)=1=\exp\left[\Psi(z)\right], \eea for all
$z$. Hence $\Psi \equiv 0$.


An operator-theoretic proof of the asymptotic independence which
also provide the rate of convergence to the factored Tracy-Widom
distributions can be found in \cite{folk}.

\setcounter{equation}{0}
\section{Studies of Laguerre Unitary Ensembles}\label{study of LUE}
This section is devoted to the study of LUE. Hence, it is understood
that all the notations $h_n(a,b)$, $\alpha_n$, $\beta_n$,
$\textsf{p}_1(n,a,b)$, etc. in this section are now associated with
$P_n(z,a,b)$ defined in \eqref{def:pn} with
$w(x)=x^{\alpha}\re^{-x}$ and $I=(0,\infty)$. We will apply a
similar theme to the LUE case as in the GUE case.

By applying lemma \ref{Lem:Ladder relation} to the Laguerre weight
$w(x)=x^{\alpha}\re^{-x}$, it is readily seen that
\begin{align}
A_n(z,a,b) & = \frac{R_n}{z}+\frac{R_{n,a}}{z-a} +
\frac{R_{n,b}}{z-b}, \label{an-new}\\
B_n(z,a,b) & = \frac{r_n}{z}+\frac{r_{n,a}}{z-a} +
\frac{r_{n,b}}{z-b}, \label{bn-new}
\end{align}
where
\begin{align}
R_n & = \frac{\alpha}{h_n(a,b)} \int_{a}^b
P_n^2(y,a,b) y^{\alpha-1}\re^{-y} dy, \label{Rn-def}\\
R_{n,a} & = a^{\alpha}\re^{-a}\frac{P_n^2(a,a,b)}{h_n(a,b)}, \label{Rna-def} \\
R_{n,b} & = -b^{\alpha}\re^{-b}\frac{P_n^2(b,a,b)}{h_n(a,b)}, \label{Rnb-def} \\
r_n & = \frac{\alpha}{h_{n-1}(a,b)} \int_{a}^b
P_n(y,a,b)P_{n-1}(y,a,b) y^{\alpha-1}\re^{-y} dy, \label{rn-def}\\
r_{n,a} & = a^{\alpha}\re^{-a}\frac{P_n(a,a,b)P_{n-1}(a,a,b)}
{h_{n-1}(a,b)}, \label{rna-def} \\
r_{n,b} & =
-b^{\alpha}\re^{-b}\frac{P_n(b,a,b)P_{n-1}(b,a,b)}{h_{n-1}(a,b)}.
\label{rnb-def}
\end{align}

Substituting the above formulas into $(S_1)$, we obtain
\begin{align}
\label{ex1} R_n+R_{n,a}+R_{n,b} &= 1,
\\
\label{ex2} r_n+r_{n+1}&=\alpha-\alpha_n R_n,
\\
\label{ex3} r_{n,a}+r_{n+1,a}&=(a-\alpha_n)R_{n,a},
\\
\label{ex4} r_{n,b}+r_{n+1,b}&=(b-\alpha_n)R_{n,b}.
\end{align}

From $(S_2')$, we find,
\begin{align}
\label{ex5} r_n^2-\alpha r_n & =\bt_nR_nR_{n-1},
\\
\label{ex6} r_{n,a}^2 & =\beta_n R_{n-1,a}R_{n,a},
\\
\label{ex7} r_{n,b}^2 & =\beta_n R_{n-1,b}R_{n,b},
\end{align}
\begin{align}
\label{ex8} &-2\frac{r_nr_{n,a}}{a}-2\frac{r_n
r_{n,b}}{b}+r_n+\alpha\frac{r_{n,a}}{a}
+\alpha\frac{r_{n,b}}{b}+\sum_{j=0}^{n-1}R_j
\nonumber \\
&=-\beta_n\Big(\frac{R_{n-1,a}R_n+R_{n-1}R_{n,a}}{a}
+\frac{R_{n-1,b}R_n+R_{n-1}R_{n,b}}{b} \Big),
\end{align}
\begin{align}
\label{ex9} &2\frac{r_nr_{n,a}}{a}+2\frac{r_{n,a}
r_{n,b}}{a-b}+r_{n,a}-\alpha\frac{r_{n,a}}{a}
+\sum_{j=0}^{n-1}R_{j,a}
\nonumber \\
&=\beta_n\Big(\frac{R_{n-1,a}R_n+R_{n-1}R_{n,a}}{a}
+\frac{R_{n-1,a}R_{n,b}+R_{n-1,b}R_{n,a}}{a-b} \Big),
\end{align}
\begin{align}
\label{ex10} &2\frac{r_nr_{n,b}}{b}+2\frac{r_{n,a}
r_{n,b}}{b-a}+r_{n,b}-\alpha\frac{r_{n,b}}{b}
+\sum_{j=0}^{n-1}R_{j,b}
\nonumber \\
&=\beta_n\Big(\frac{R_{n-1,b}R_n+R_{n-1}R_{n,b}}{b}
+\frac{R_{n-1,a}R_{n,b}+R_{n-1,b}R_{n,a}}{b-a} \Big).
\end{align}
From $(S_2)$, we find
\begin{align}
\label{ex11} 1+r_{n+1}-r_{n}+r_{n+1,a}-r_{n,a}+r_{n+1,b}-r_{n,b}&=0, \\
\label{ex12}\alpha_n(r_n-r_{n+1})&=\beta_{n+1}R_{n+1}-\beta_nR_{n-1}, \\
\label{ex13}(a-\alpha_n)(r_{n+1,a}-r_{n,a})&=\beta_{n+1}R_{n+1,a}-\beta_nR_{n-1,a}, \\
\label{ex14}(b-\alpha_n)(r_{n+1,b}-r_{n,b})&=\beta_{n+1}R_{n+1,b}-\beta_nR_{n-1,b}.
\end{align}

The sum of \eqref{ex8}--\eqref{ex10} yields
\begin{equation}\label{sum of r}
r_n+r_{n,a}+r_{n,b}+n=0,
\end{equation}
where we have made use of \eqref{ex1}. This equation can also be
obtained by a telescopic sum of \eqref{ex11}. Summing
\eqref{ex2}--\eqref{ex4}, we see from \eqref{ex1} and \eqref{sum of
r} that
\begin{equation}\label{alphan and Rn}
\alpha_n=\alpha+aR_{n,a}+bR_{n,b}+2n+1.
\end{equation}
By \eqref{ex9}--\eqref{ex10}, it is easily seen that
\begin{align}
&a\sum_{j=0}^{n-1}R_{j,a}+b\sum_{j=1}^{n-1}R_{j,b}
\nonumber \\
&=\beta_n\Big((R_{n-1,a}+R_{n-1,b})R_n+(R_{n,a}+R_{n,b})R_{n-1}+
R_{n-1,a}R_{n,b}+R_{n-1,b}R_{n,a}\Big) \nonumber \\
&~~~-2r_n(r_{n,a}+r_{n,b})+(\alpha-a)r_{n,a}+(\alpha-b)r_{n,b}-2r_{n,a}r_{n,b}.
\end{align}
Now, we use \eqref{ex1} and \eqref{sum of r} to eliminate $r_n$ and
$R_n$ ($R_{n-1}$), \eqref{ex6} and \eqref{ex7} to eliminate
$R_{n-1,a}$ and $R_{n-1,b}$ in the above equation, it follows that
\begin{align}\label{asumRa+bsumRb}
a\sum_{j=0}^{n-1}R_{j,a}+b\sum_{j=0}^{n-1}R_{j,b}
&=\beta_n(R_{n,a}+R_{n,b})+\frac{r_{n,a}^2}{R_{n,a}}(1-R_{n,b})+
\frac{r_{n,b}^2}{R_{n,b}}(1-R_{n,a})
\nonumber \\
&~~~+(2n+\alpha-a)r_{n,a}+(2n+\alpha-b)r_{n,b}+2r_{n,a}r_{n,b}.
\end{align}

\subsection{Toda evolution}
As in the GUE case, it is easily verified that, in the present case,
we still have
\begin{equation}\label{partial log hn}
\partial_a \ln h_n=-R_{n,a}, \qquad
\partial_b \ln h_n=-R_{n,b}.
\end{equation}
\begin{equation}
\label{partial ln betan}
\partial_a \ln \beta_n = R_{n-1,a}-R_{n,a}, \qquad
\partial_b \ln \beta_n = R_{n-1,b}-R_{n,b},
\end{equation}
with $R_{n,a}$ and $R_{n,b}$ defined in \eqref{Rna-def} and
\eqref{Rnb-def}, respectively, and
\begin{equation}\label{partial p1}
\partial_a \textsf{p}_1(n,a,b)=r_{n,a},
\qquad \partial_b \textsf{p}_1(n,a,b)=r_{n,b}.
\end{equation}
\begin{equation}\label{partial alphan}
\partial_a \alpha_n = r_{n,a}-r_{n+1,a}, \qquad
\partial_b \alpha_n = r_{n,b}-r_{n+1,b},
\end{equation}
where $r_{n,a}$ and $r_{n,b}$ are given in \eqref{rna-def} and
\eqref{rnb-def}, respectively.

With the above preparations, we are ready to state a lemma which
gives a two variable version of the Toda equation for the recurrence
coefficients $\alpha_n$ and $\beta_n$ in the present case:
\begin{lemma}
For LUE, we have
\begin{align}
\label{betan and alphan}
(a\partial_a+b\partial_b)\beta_n&=\beta_n(\alpha_{n-1}-\alpha_n+2),
\\
\label{alphan and betan}
(a\partial_a+b\partial_b-1)\alpha_n&=\beta_n-\beta_{n+1}.
\end{align}
\end{lemma}
\begin{proof}
From \eqref{partial ln betan}, it is easily seen that
\begin{equation}
\frac{(a\partial_a+b\partial_b)\beta_n}{\beta_n}=aR_{n-1,a}+bR_{n-1,b}
-(aR_{n,a}+bR_{n,b}).
\end{equation}
Taking into account of \eqref{alphan and Rn}, this gives
\begin{equation}
\frac{(a\partial_a+b\partial_b)\beta_n}{\beta_n}=\alpha_{n-1}-\alpha_n+2,
\end{equation}
which is \eqref{betan and alphan}.

To establish \eqref{alphan and betan}, we make the following
decomposition
\begin{align}\label{decomp}
(a\partial_a+b\partial_b)\alpha_n
&=\big((a-\alpha_n)\partial_a+(b-\alpha_n)\partial_b\big)\alpha_n
+\alpha_n(\partial_a+\partial_b)\alpha_n.
\end{align}
By \eqref{partial alphan} and \eqref{sum of r}, we have
\begin{align}\label{part2}
\alpha_n(\partial_a+\partial_b)\alpha_n
=\alpha_n(r_{n,a}+r_{n,b}-r_{n+1,a}-r_{n+1,b})=\alpha_n(1+r_{n+1}-r_n).
\end{align}
On the other hand, it follows from \eqref{partial alphan},
\eqref{ex13}, \eqref{ex14} and \eqref{ex1} that
\begin{align}\label{part1}
&\big((a-\alpha_n)\partial_a+(b-\alpha_n)\partial_b\big)\alpha_n
\nonumber \\
&=(a-\alpha_n)(r_{n,a}-r_{n+1,a})+(b-\alpha_n)(r_{n,b}-r_{n+1,b})
\nonumber \\
&=\beta_n(R_{n-1,a}+R_{n-1,b})-\beta_{n+1}(R_{n+1,a}+R_{n+1,b})
\nonumber \\
&=\beta_n-\beta_{n+1}+\beta_{n+1}R_{n+1}-\beta_nR_{n-1}
\nonumber \\
&=\beta_n-\beta_{n+1}+\alpha_n(r_n-r_{n+1}),
\end{align}
where we also make use of \eqref{ex12} in the last step of the above
equation. Substituting \eqref{part1} and \eqref{part2} into
\eqref{decomp} gives us \eqref{alphan and betan}.
\end{proof}

\subsection{Derivation of partial differential equation}

We set
\begin{equation}\label{Hn}
H_n(a,b):=(a\partial_a+b\partial_b)\ln D_n(a,b),
\end{equation}
where $D_n$ is the Hankel determinant defined in \eqref{Hankel det}.
It is the aim of this section to derive a pde satisfied by $H_n$.

We note that, the equations \eqref{part a ln Dn} and \eqref{part b
ln Dn} still hold in the present case, it then follows from
\eqref{asumRa+bsumRb} that
\begin{align}\label{Hn rep 1}
H_n&=-a\sum_{j=0}^{n-1} R_{j,a}-b\sum_{j=0}^{n-1} R_{j,b} \nonumber \\
&=(a-\alpha-2n)r_{n,a}+(b-\alpha-2n)r_{n,b}-2r_{n,a}r_{n,b}
\nonumber \\
&~~~-\beta_n(R_{n,a}+R_{n,b})-\frac{r_{n,a}^2}{R_{n,a}}(1-R_{n,b})-
\frac{r_{n,b}^2}{R_{n,b}}(1-R_{n,a}).
\end{align}
In view of \eqref{alphan and Rn} and \eqref{alphan representation},
we also have
\begin{align}\label{Hn rep 2}
H_n&=-a\sum_{j=0}^{n-1} R_{j,a}-b\sum_{j=0}^{n-1} R_{j,b} \nonumber
\\
&=\sum_{j=0}^{n-1}(\alpha+2j+1-\alpha_j(a,b))=n(\alpha+n)+\textsf{p}_1(n,a,b).
\end{align}
This, together with \eqref{partial p1}, implies
\begin{equation}\label{partial Hn}
\partial_a H_n=r_{n,a},\qquad \partial_b H_n=r_{n,b}.
\end{equation}

Next, we derive representations of $\beta_n$, $R_{n,a}$ and
$R_{n,b}$ in terms of $H_n$ and its partial derivatives. To this
end, we use \eqref{sum of r} and \eqref{ex1} to eliminate $r_n$ and
$R_n$ in \eqref{ex5}, and then use \eqref{ex6} and \eqref{ex7} to
eliminate the resulting $R_{n-1,a}$ and $R_{n-1,b}$, it follows that
\begin{align}
&-\beta_n(R_{n,a}+R_{n,b})-\frac{r_{n,a}^2}{R_{n,a}}(1-R_{n,b})-
\frac{r_{n,b}^2}{R_{n,b}}(1-R_{n,a}) \nonumber \\
&=-\beta_n+(n+r_{n,a}+r_{n,b})^2+\alpha(n+r_{n,a}+r_{n,b})-r_{n,a}^2-r_{n,b}^2.
\end{align}
Inserting the above equation into \eqref{Hn rep 1}, we obtain after
some simplification that
\begin{equation}
\beta_n=n(n+\alpha)-H_n+ar_{n,a}+br_{n,b},
\end{equation}
or equivalently, taking into account of \eqref{partial Hn},
\begin{equation}\label{betn in Hn}
\beta_n=n^2+\alpha n-H_n+a\partial_aH_n+b\partial_bH_n.
\end{equation}
From \eqref{partial ln betan}, \eqref{ex6} and \eqref{ex7}, we
further have
\begin{equation}\label{partial betan}
\partial_a\beta_n=\frac{r_{n,a}^2}{R_{n,a}}-\beta_nR_{n,a},\qquad
\partial_b\beta_n=\frac{r_{n,b}^2}{R_{n,b}}-\beta_nR_{n,b}.
\end{equation}
Using \eqref{betn in Hn} and \eqref{partial Hn} in \eqref{partial
betan}, it is readily seen that
\begin{align}
a\partial_a^2H_n+b\partial_a\partial_bH_n&=\frac{(\partial_aH_n)^2}{R_{n,a}}
-(n^2+\alpha n-H_n+a\partial_aH_n+b\partial_bH_n)R_{n,a}, \\
b\partial_b^2H_n+a\partial_a\partial_bH_n&=\frac{(\partial_bH_n)^2}{R_{n,b}}
-(n^2+\alpha n-H_n+a\partial_aH_n+b\partial_bH_n)R_{n,b}.
\end{align}
Solving the above quadratic equations for $R_{n,a}$ and $R_{n,b}$,
we obtain
\begin{align}
\label{Rna in Hn}
R_{n,a}&=\frac{-(a\partial_a^2H_n+b\partial_a\partial_bH_n)\pm
\sqrt{\Delta_1}}{2(n^2+\alpha n-H_n+a\partial_aH_n+b\partial_bH_n)},
\\
\label{Rnb in Hn}
R_{n,b}&=\frac{-(b\partial_b^2H_n+a\partial_a\partial_bH_n)\pm
\sqrt{\Delta_2}}{2(n^2+\alpha n-H_n+a\partial_aH_n+b\partial_bH_n)},
\end{align}
where
\begin{equation}\label{Delta1}
\Delta_1=(a\partial_a^2H_n+b\partial_a\partial_bH_n)^2+4(\partial_aH_n)^2
(n^2+\alpha n-H_n+a\partial_aH_n+b\partial_bH_n)
\end{equation}
and
\begin{equation}\label{Delta2}
\Delta_2=(b\partial_b^2H_n+a\partial_a\partial_bH_n)^2+4(\partial_bH_n)^2
(n^2+\alpha n-H_n+a\partial_aH_n+b\partial_bH_n).
\end{equation}

Finally, substituting \eqref{partial Hn}, \eqref{betn in Hn},
\eqref{Rna in Hn} and \eqref{Rnb in Hn} into \eqref{Hn rep 1} yields
\begin{align}\label{pde for Hn}
&2(n^2+\alpha n-H_n+a\partial_aH_n+b\partial_bH_n)(H_n
+(2n+\alpha-a)\partial_aH_n+(2n+\alpha-b)\partial_bH_n
+2\partial_aH_n\partial_bH_n) \nonumber \\
&+(a\partial_a^2H_n+b\partial_a\partial_bH_n)
(b\partial_b^2H_n+a\partial_a\partial_bH_n)\nonumber \\
&=2(n^2+\alpha
n-H_n+a\partial_aH_n+b\partial_bH_n)(\pm\sqrt{\Delta_1}\pm\sqrt{\Delta_2})
\pm\sqrt{\Delta_1\Delta_2},
\end{align}
where $\Delta_i$, $i=1,2$ is given in \eqref{Delta1} and
\eqref{Delta2}, respectively. Denote by
\begin{align*}
k&:=2(n^2+\alpha n-H_n+a\partial_aH_n+b\partial_bH_n)(H_n
+(2n+\alpha-a)\partial_aH_n+(2n+\alpha-b)\partial_bH_n
\\
&~~~~+2\partial_aH_n\partial_bH_n)
+(a\partial_a^2H_n+b\partial_a\partial_bH_n)
(b\partial_b^2H_n+a\partial_a\partial_bH_n)
\end{align*}
and
\begin{equation*}
l:=2(n^2+\alpha n-H_n+a\partial_aH_n+b\partial_bH_n),
\end{equation*}
we can rewrite equation \eqref{pde for Hn} in the following
equivalent form:
\begin{align}\label{final pde LUE}
&\Big(\big((k^2-l^2(\Delta_1+\Delta_2)-\Delta_1\Delta_2)^2-4l^2\Delta_1\Delta_2
(l^2+\sqrt{\Delta_1\Delta_2}+\Delta_1+\Delta_2)\big)^2 \nonumber \\
&-16l^6(\Delta_1\Delta_2)^2
(\Delta_1+\Delta_2)\Big)^2-1024l^{12}(\Delta_1\Delta_2)^5=0.
\end{align}

Suppose there is no $a$-dependence in $H_n$, the equation
\eqref{final pde LUE} reduces to
\begin{equation}\label{no a dependence}
(b\partial_b^2H_n)^2
=(H_n+(2n+\alpha-b)\partial_bH_n)^2-4(\partial_bH_n)^2 (n^2+\alpha
n-H_n+b\partial_bH_n).
\end{equation}
The equation \eqref{no a dependence} is nothing but the
Okamoto-Jimbo-Miwa $\sigma$-form of the Painlev\'{e} V equation
\cite{JM}:
\begin{equation}
\begin{aligned}
(z\sigma'')^2&=\big(\sigma-z\sigma'+2(\sigma')^2
+(\nu_0+\nu_1+\nu_2+\nu_3)\sigma'\big)^2 \\
&~~~-4(\sigma'+\nu_0)(\sigma'+\nu_1)(\sigma'+\nu_2)(\sigma'+\nu_3),
\end{aligned}
\end{equation}
with
\begin{equation}
\nu_0=n,\qquad \nu_1=n+\alpha,\qquad \nu_2=\nu_3=0.
\end{equation}
We have the same conclusion if there is no $b$-independence in
$H_n$.


\subsection{Scaling of PDE }
In this section, we will scale the pde obtained in \eqref{pde for
Hn} and show that its solution is asymptotically equal to the sum of
Tracy-Widom left and right distributions under certain delicate
scaling.

For this purpose, we set
\begin{equation}
\alpha=\beta n, \qquad \beta>0,
\end{equation}
and denote by
\begin{equation}
L:=2+\beta-2\sqrt{2+\beta},\qquad R:=2+\beta+2\sqrt{2+\beta}.
\end{equation}
Note that $LR=\beta^2$. We then scale the left and right soft edges
$a$ and $b$ as follows:
\begin{equation}\label{change of var LUE}
\begin{aligned}
a=Ln+cL^{2/3}n^{1/3}x, \qquad b=Rn+cR^{2/3}n^{1/3}y,
\end{aligned}
\end{equation}
with $c>0$. Clearly, it is easily seen that
\begin{equation}
\partial_a=\frac{\partial_x}{cL^{2/3}n^{1/3}},\qquad
\partial_b=\frac{\partial_y}{cR^{2/3}n^{1/3}},
\end{equation}
and
\begin{equation}
a\partial_a=\left(\frac{L^{1/3}}{c}n^{2/3}+x\right)\partial_x,\qquad
b\partial_b=\left(\frac{R^{1/3}}{c}n^{2/3}+y\right)\partial_y.
\end{equation}
Recall
\begin{equation}
H_n(a,b)=(a\partial_a+b\partial_b)\ln
D_n(a,b)=(a\partial_a+b\partial_b)\ln \textsf{Prob}(n,a,b),
\end{equation}
and in the new variables $x$, $y$ this becomes
\begin{equation}
\begin{aligned}
&H_n(Ln+cL^{2/3}n^{1/3}x,Rn+cR^{2/3}n^{1/3}y) \\
&=\left[\left(\frac{L^{1/3}}{c}n^{2/3}+x\right)\partial_x
+\left(\frac{R^{1/3}}{c}n^{2/3}+y\right)\partial_y\right]\ln D_n.
\end{aligned}
\end{equation}
Let
\begin{equation}
\tH(x,y,n):=\frac{1}{n^{2/3}}H_n(Ln+cL^{2/3}n^{1/3}x,Rn+cR^{2/3}n^{1/3}y).
\end{equation}
After substituting the change of variables \eqref{change of var
LUE}, the leading term of \eqref{pde for Hn} is of order $n^{8/3}$
and produces the following limiting pde:
\begin{align}\label{limiting pde LUE}
&4c^3\sqrt{1+\beta}\beta^{4/3}\px\tH\py\tH(\tH-x\px\tH-y\py\tH)\nonumber \\
&+\px\tH(L^{4/3}(\px\py\tH)^2+\beta^{4/3}(\py^2\tH)^2+2\beta^{2/3}L^{2/3}\px\py\tH\py^2\tH
+4c\beta^{2/3}L^{1/3}(\py\tH)^3)\nonumber \\
&-\py\tH(R^{4/3}(\px\py\tH)^2+\beta^{4/3}(\px^2\tH)^2+2\beta^{2/3}R^{2/3}\px\py\tH\px^2\tH
+4c\beta^{2/3}R^{1/3}(\px\tH)^3)=0.
\end{align}

As in GUE case, it turns out that pde \eqref{limiting pde LUE}
admits the following factorization ansatz:
\begin{equation}\label{tilde H LUE}
\tH(x,y)=f(x)+g(y),
\end{equation}
where $f$ and $g$ satisfy the $\sigma$-form of a particular
Painlev\'{e} II, respectively. More precisely,
\begin{align}
(f''(x))^2&=4c^3\sqrt{1+\beta}f(x)f'(x)-4c^3\sqrt{1+\beta}x(f'(x))^2
-\frac{4c}{L^{1/3}}(f'(x))^3,
\label{PII sigma form f}\\
(g''(y))^2&=-4c^3\sqrt{1+\beta}g(y)g'(y)+4c^3\sqrt{1+\beta}y(g'(y))^2
-\frac{4c}{R^{1/3}}(g'(y))^3.\label{PII sigma form g}
\end{align}
It is readily seen that \eqref{tilde H LUE}--\eqref{PII sigma form
g} satisfy \eqref{limiting pde LUE} identically. This proves that
\begin{equation}
\lim_{n\to\infty}\frac{1}{n^{2/3}}H_n(Ln+cL^{2/3}n^{1/3}x,Rn+cR^{2/3}n^{1/3}y)
=f(x)+g(y),
\end{equation}
where $f$ and $g$ is Tracy-Widom left and right distribution,
respectively.

{\bf Acknowledgements} We would like to thank Iain Johnstone for
bringing this problem to our attention. Lun Zhang is supported in
part by the Belgian Interuniversity Attraction Pole P06/02m and by
FWO-Flanders project G.0427.09.

\end{document}